\documentclass[lettersize,journal]{IEEEtran}
\usepackage{amsmath,amsfonts}
\usepackage{algorithmic}
\usepackage{algorithm}
\usepackage{array}
\usepackage[caption=false,font=normalsize,labelfont=sf,textfont=sf]{subfig}
\usepackage{textcomp}
\usepackage{stfloats}
\usepackage{url}
\usepackage{verbatim}
\usepackage{graphicx}
\usepackage{cite}
\usepackage{xcolor}
\usepackage{soul}
\usepackage{tikz}
\usepackage[utf8x]{inputenc}
\DeclareMathOperator*{\E}{\mathbb{E}}
\newcommand{\probP}{\text{I\kern-0.15em P}}
\newcommand{\ubar}[1]{\text{\b{$#1$}}}
\begin{document}

\title{Risk-Aware Congestion Management with Capacity Limitation Contracts and Redispatch}

\author{Bart van der Holst,~\IEEEmembership{Student Member,~IEEE, Phuong Nguyen,\\~\IEEEmembership{Member,~IEEE}, Johan Morren,~\IEEEmembership{Member,~IEEE}, Koen Kok,~\IEEEmembership{Member,~IEEE}}
\thanks{The Netherlands Enterprise Agency financially supported
this work through the GO-E project (MOOI32001).}
\thanks{All authors are with the Eindhoven University of Technology, Department of Electrical Engineering, Eindhoven, 5612 AP. Johan Morren is also with Enexis Netbeheer, Asset Management, 's Hertogenbosch, 5223 MB.}}




\newcommand\copyrighttext{%
  \footnotesize \textcopyright 2025 IEEE. Personal use of this material is permitted.
  Permission from IEEE must be obtained for all other uses, in any current or future
  media, including reprinting/republishing this material for advertising or promotional
  purposes, creating new collective works, for resale or redistribution to servers or
  lists, or reuse of any copyrighted component of this work in other works.}
\newcommand\copyrightnotice{%
\begin{tikzpicture}[remember picture,overlay]
\node[anchor=south,yshift=10pt] at (current page.south) 
  {\fbox{\parbox{\dimexpr\textwidth-\fboxsep-\fboxrule\relax}{\copyrighttext}}};
\end{tikzpicture}%
}

\maketitle
\copyrightnotice

\begin{abstract}
This paper presents the coordination of two congestion management instruments — capacity limitation contracts (CLCs) and redispatch contracts (RCs) — as a risk-aware resource allocation problem. We propose that the advantages and drawbacks of these instruments can be represented as operational risk profiles and can be balanced through coordination. To this end, we develop a chance-constrained two-stage stochastic mixed-integer program for a system operator procuring flexibility from an aggregator managing a fleet of electric vehicles (EVs). The model captures uncertainty in EV charging and redispatch market conditions, using real order book data from the Dutch redispatch market (GOPACS).
Results indicate that combining CLCs and RCs is generally the most cost-effective approach to mitigate risks associated with each instrument, but the optimal mix depends on fleet size and RC activation timing. Large uncertainty about EV loading increases RC activation intraday to correct for forecasting errors at the earlier CLC stage. For large fleet sizes (e.g. 25.000) the optimal policy limits redispatch due to market liquidity risks in the immature redispatch market. This risk increases for later redispatch activation due to shrinking trading windows for redispatch products. These findings highlight how various sources of uncertainty can impact the optimal trade-off between congestion management instruments.


\end{abstract}


\section{Introduction}
\label{sec: introduction}
\subsection{Motivation}
\label{sec: motivation}
Traditionally, System Operators (SOs) have relied on grid reinforcement to prevent congestion, but several countries are now adopting congestion management instruments as an additional approach \cite{CPUC_D20_09_035}\cite{givisiez2023assessing}\cite{CIRED}. These instruments — such as network tariffs, connection agreements, market-based approaches, and direct control — can coexist in practice, raising questions on potential synergies and incompatibilities between them \cite{ormeno2024unlocking}.

In 2022, capacity limitation contracts (CLCs) and redispatch contracts (RCs) were introduced to both Distribution System Operators (DSOs) and the Transmission System Operator (TSO) in the Netherlands \cite{Netcode}. These two bilateral contracts can be used by SOs to procure flexibility from Congestion Service Providers (CSPs). The CLC is a capacity-based product that limits a CSP’s portfolio connection capacity and can be activated a few hours before the day-ahead (DA) market closes. In contrast, the RC is an energy-based product, enforcing CSP participation in the Dutch redispatch market (GOPACS) and can be activated continuously intraday.

For day-to-day congestion management, the SOs thus need to decide how much capacity and volume to procure via the CLCs and RCs, respectively. We emphasize that, although this decision problem may initially seem specific to Dutch SOs, the setting of a capacity-based product for day-ahead congestion management and a redispatch product for the intraday is quite generic. For instance, non-firm connection agreements are very similar to CLCs, and have been introduced in countries like France, Belgium and the UK \cite{CEER}. Moreover, many other countries operate redispatch markets at the transmission level, and some, like the UK, are also facilitating intraday redispatch at the distribution level \cite{flexiblepower}. In other words, though this paper focuses on the congestion management instruments from the Netherlands, the contributions of this paper have a broader applicability.

In our previous work, we introduced a decision framework for congestion management with CLCs and RCs based solely on costs \cite{van2024activation}. In this paper, we take a more complete approach based on the work of Hennig \textit{et al.} \cite{hennig2023congestion}, who characterized several congestion management instruments by operational risk profiles. In this work, we expand on this idea for congestion management with multiple instruments by characterizing the risk profiles of the CLC and RC and presenting their daily activation as a risk-aware resource allocation problem over a portfolio of congestion management instruments.

\subsection{Related Work}
This section reviews relevant literature on congestion management that incorporates a risk-aware approach and/or multiple instruments. Various studies are available on risk-aware congestion management, but almost all focus on a single congestion management instrument. Within the distribution system, these efforts focus on locational marginal pricing \cite{amanbek2022distribution}\cite{liu2016distribution}\cite{ni2017congestion}, dynamic tariffs \cite{huang2016uncertainty}\cite{huang2018dynamic}\cite{shen2022robust}, and local flexibility markets \cite{esmat2018decentralized}\cite{paredes2023uncertainty}. For the transmission system, most of the work is naturally on stochastic market-clearing algorithms for DA markets \cite{bjorndal2016congestion}\cite{wu2019chance}\cite{beyzaee2020risk}, and/or redispatch markets \cite{hojjat2013probabilistic}\cite{liu2015two}. Two studies worth highlighting are from Hennig \textit{et al.} \cite{hennig2024risk} and Panda \textit{et al.} \cite{panda2024quantifying}. The former compares three capacity-based congestion management instruments, one of them being the CLC addressed in this study. The key difference between the considered instruments is the lead time the SO announces capacity limitations (long-term, day-ahead, or real-time). The authors examine how this timing affects operational risk for both the SO and contracted parties \cite{hennig2024risk}. Panda \textit{et al.} train a machine learning model on EV charging data to predict the likelihood that congestion management can be provided via CLCs and RCs \cite{panda2024quantifying}. Both papers thus also focus on the instruments from the Netherlands, but only consider one instrument at a time, not focusing on trade-offs between multiple instruments.

Examples of studies on stacking multiple congestion management instruments come from Chen \textit{et al.} \cite{shen2018comprehensive}\cite{shen2022coordination}. The first work combines dynamic tariffs, grid reconfiguration, and a redispatch product sequentially \cite{shen2018comprehensive}, whereas the second study presents a more coordinated approach for the optimal use of a dynamic tariff and the redispatch product together. Similar bilevel frameworks were presented in \cite{pediaditis2022synergies}\cite{pediaditis2024integrating}\cite{fattaheian2020incentive}. The first two combine a dynamic tariff with a capacity-based product, whereas the third coordinates grid reconfiguration with redispatch. However, none of these studies considered uncertainty in their approach, thereby overlooking the risks associated with the congestion management instruments. Only \cite{ansari2021congestion} considers risk in a limited way, providing a two-step approach in which a DSO first performs a load flow to estimate the expected congestion and decides on a "risk level", based on the predicted magnitude of the problem. Depending on the risk level, either a redispatch instrument or a dynamic tariff is applied.

To the authors' knowledge, the literature does not present another risk-aware coordination framework for multiple congestion management instruments. That being said, the resulting problem structure shows similarities with other problems in the electricity market literature. From a modeling point of view, these include: 1) finding a trading strategy to arbitrage between the DA and intraday markets \cite{meese2018optimized}\cite{demir2023statistical}\cite{shinde2022modified}, and 2) hedging financial risk caused by capacity shortages on interconnections \cite{androcec2006different}\cite{androcec2006different}. In terms of methodology, our work shares the most similarities with these studies.

\subsection{Contributions}
To address the established gap in the literature, this study makes the following contributions:
\begin{itemize}
 \item Formulating congestion management with CLCs and RCs as a risk-aware resource allocation problem, resulting in a chance-constrained two-stage mixed-integer linear stochastic program with recourse.
\item Providing the first analysis of the order book of the Dutch national redispatch market GOPACS to include price and liquidity risks associated with the RCs.
 \item Analyzing the effect of EV fleet sizes and the timing of activating redispatch on the optimal allocation of costs over CLCs and RCs. 
\end{itemize}

\section{Congestion Management Instruments and Their Risks Profile}
\label{sec: products}
Since 2022, Dutch DSOs and the TSO have access to two types of contracts for congestion management: capacity limitation contracts and redispatch contracts \cite{Netcode}. Both contracts have specific characteristics and associated operational risks from the perspective of the SO, CSPs, and connected parties. We will now discuss these contracts and their characteristics in more detail.

\subsection{Capacity Limitation Constracts}
\label{sec: CLC}
\begin{figure}[!t]
\centering
\includegraphics[width=1.05\linewidth]{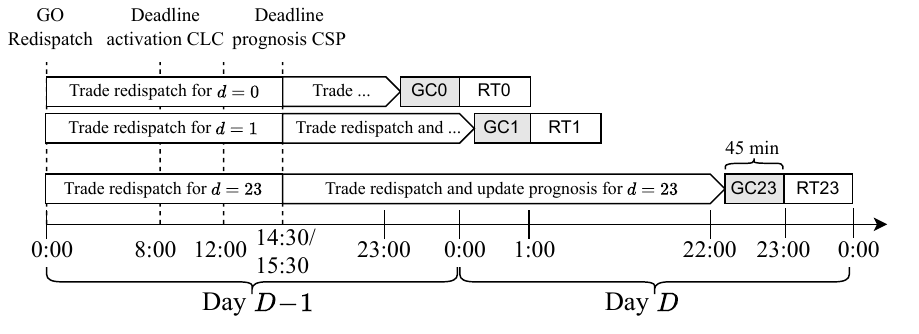}
\caption{Timeline with important events for congestion management with capacity limitation contracts (CLC) and redispatch contracts (RCs). Other abbreviations are: GO: gate opening, GC: gate closure, RT: real-time (moment of delivery), CSP: congestion service provider.}
\label{fig: timeline}
\end{figure}

A CLC grants the SO the right to reduce the available network capacity of a connected party for the next day (day $D$). 
The minimum available capacity is specified in the CLC, together with a compensation scheme. 
In practice, most SOs communicate contract activation at 8:00 $D\!-\!1$ at the latest, to give CSPs the time to adjust their schedules before the DA market closes. This moment of activation is also shown in Fig. \ref{fig: timeline}, which presents a timeline of all relevant steps for both CLC and RC activation. 

Depending on the compensation scheme negotiated in the CLC, either the SO or CSP suffers from price risk. If, for example, the CLC sets a fixed price per unit of reduced capacity over a time period (e.g., 15 or 60 minutes), there is no price risk for the SO for the for the contract activation. From the perspective of the CSP, this agreed price may turn out to be (un)favorable compared to prices in e.g. the DA market on day $D$. An alternative is to compensate the CSP based on missed income from the DA market on $D$. This shifts the price risk towards the SO. However, even in that case the CSP can face opportunity risks, since the capacity limitation may restrict participation in reserve markets for e.g. Frequency Containment Reserves or Frequency Restoration Reserves. Capacity limitations can also hinder intraday trading and implicit balancing.

For the SO, uncertainty about loads and generation on day $D$ results in several risks. Predicting intermittent load and generation 16 to 40 hours in advance can be challenging. If not sufficient capacity limitation is procured, the SO risks ending up with residual congestion. If too much capacity reduction is procured, the SO risks overpaying. Additionally, if the flexibility of the CSP is overestimated, asset owners in the portfolio of the CSP risk being curtailed.

\subsection{Redispatch Contracts}
\label{sec: RC}
Redispatch refers to adjusting the amount of energy consumed or generated relative to a baseline after the DA market has closed. Redispatch can be contracted using an RC, which provides more security for the SO than depending on free redispatch offers.

In 2019, the GOPACS platform was introduced to host a shared national redispatch market for both Dutch DSOs and TenneT \cite{GOPACS}. GOPACS hosts a continuous market with pay-as-bid pricing. When an SO expects congestion in a region, it can activate an RC there to force a CSP to submit an order to provide upward/downward flexibility for some moment(s) of delivery on day $D$. In contrast to orders in the intraday market, this bid contains locational information about the connection providing the flexibility, and it can only be matched with a counter-bid from outside the congested area, submitted by a third party. In this way, the power flow through the congested asset is reduced, and the total balance of the system is not affected by activating the RC. All orders submitted in GOPACS are limit orders, which means that partial matching is possible and that the unmatched volume remains in the order book for a future trade. There typically exists a price gap between the bid of the CSP and the counter-volume in the order book. In that case, the SO pays the spread to the seller, which introduces costs for the SO. Due to the continuous nature of the redispatch market, redispatch products for a specific time of delivery on day $D$ can be bought and sold continuously, up to gate closure (GC), 45 minutes before delivery. The baseline used for redispatch is called the \textit{prognosis}, which the CSPs must submit to the SO at either 14:30 $D-1$ (distribution system) or 15:30 $D-1$ (transmission system) \cite{Netcode}. These events are also presented in the timeline in Fig. \ref{fig: timeline}.

In terms of the RC risk profile, there is clearly both price risk and liquidity risk from the perspective of the SO. Although the SO can use the RC to contractually arrange bid prices with CSPs in the congested area, the price spread with offers from counter bids is highly uncertain and has shown to be high in the past (see Section \ref{sec: redispatch market}). Additionally, there is liquidity risk as it is uncertain whether enough counter-volume will be available for a product, especially if an RC is activated close to delivery. In that case, the SO can end up with residual congestion.

The fact that redispatch requires self-reported baselines from market parties also poses several risks for the SO \cite{ziras2021baselines}. Currently, it is hard to check the reliability of a CSP's prognosis day-ahead. This creates the risk of increase-decrease gaming behavior \cite{hirth2020market}. Even if the prognosis reflects the true intentions of the CSP, inaccuracies due to forecasting errors are bound to arise, which could result in overprocurment of redispatch by the SO. However, this risk is lower than for CLCs, as the continuous nature of the redispatch market provides SOs significantly more flexibility to correct errors made in congestion forecasts. With GCs within one hour of delivery, the risk of over- or underprocurement of flexibility, and the risk of curtailing end-users can be greatly reduced.

\begin{figure}[!t]
\centering
\includegraphics[width=1.0\linewidth]{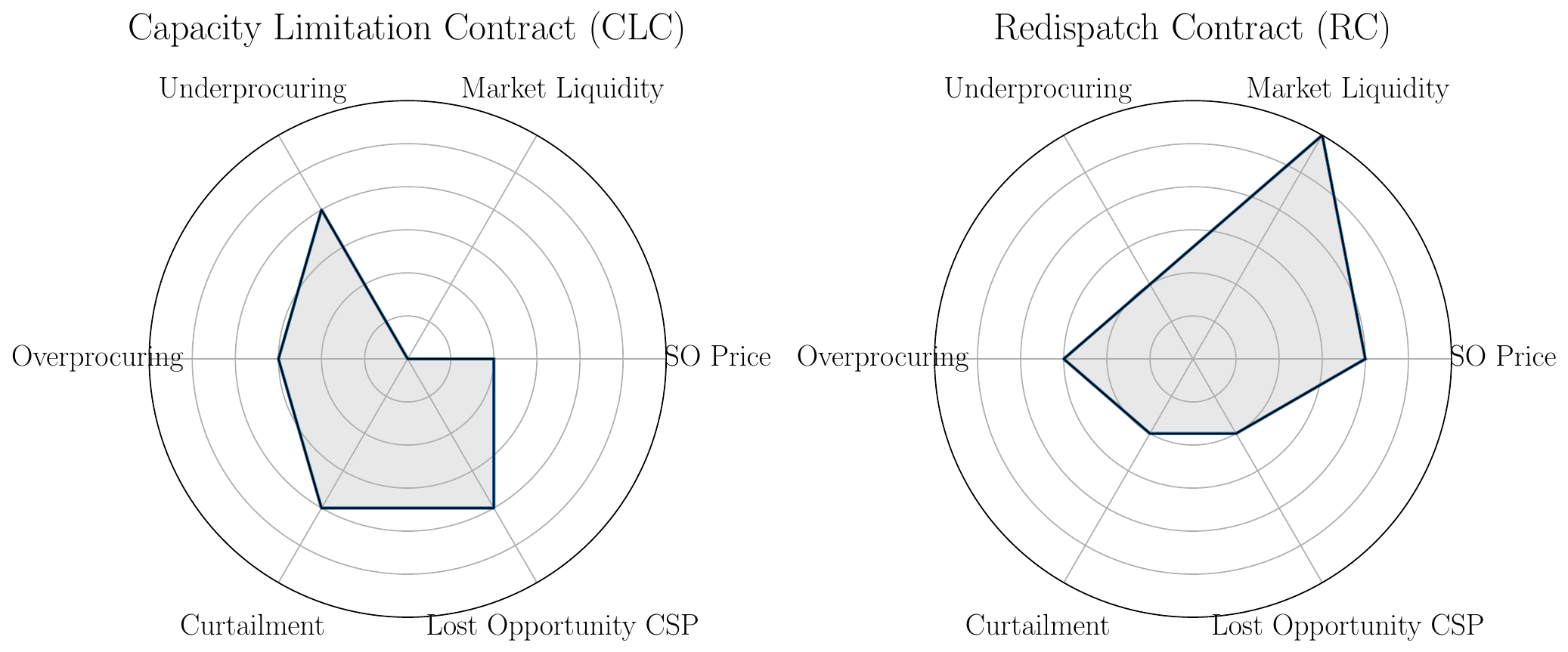}
\caption{Illustration of the risk profiles of capacity limitation contracts and redispatch contracts, evaluated across six operational risks discussed in Sections \ref{sec: CLC} and \ref{sec: RC}. Starting on the left, and moving clockwise in the figure, we have: the risk of over- or underprocuring flexibility by the SO; the risk of not finding enough flexibility in the (redispatch) market; the risk the SO pays a high price for flexibility, the risk CSPs miss out on other opportunities; and the risk that connections in the portfolio of the CSP need to be curtailed.}
\label{fig: risk profiles}
\end{figure}

To summarize the characteristics of CLC and RC, Fig. \ref{fig: risk profiles} presents a graphical representation of the risk profiles of both instruments.

\section{Decision Model}
\label{sec: decision model}
This section presents a two-stage stochastic program with recourse to capture the risk-aware coordination problem of CLC and RC activation. For simplicity, we assume that the SO has both a CLC and a RC with a single CSP in the congested area and that its portfolio consists of EVs only. Considering EVs is interesting as a case-study, as there is a great potential to use their flexibility for congestion management \cite{das2025congestion}. At the same time, the load and flexibility of an EV fleet can be difficult to predict day ahead, making a risk-aware decision model particularly relevant. The model can be extended to consider more heterogeneous portfolios, but it requires an account of how the distributions of the stochastic asset parameters develop over time. Section \ref{sec: EV fleet} details how this aspect can be approached for EV fleets in particular.

\begin{figure}[!b]
\centering
\includegraphics[width=1.05\linewidth]{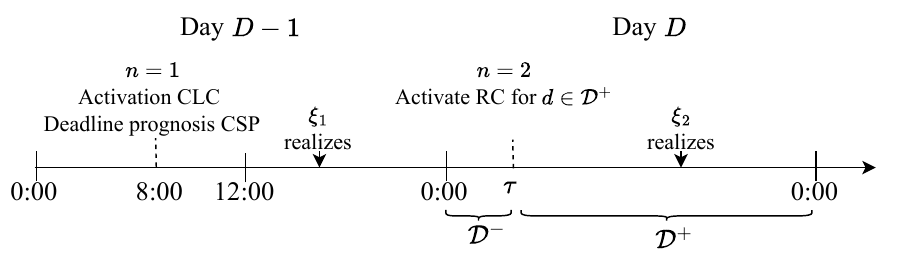}
\caption{Simplified timeline of the decision process presented in Fig. \ref{fig: timeline} into two decision stages (dashed lines) and two uncertainty stages (arrows). The second decision stage involves activating redispatch from the RCs and takes place at timestamp $\tau$. This moment can be either on day $D-1$ or day $D$. The latter case is presented in this figure.}
\label{fig: decision}
\end{figure}

\subsection{Towards a Two-Stage Stochastic Program with Recourse}
\label{sec: two-stage}
We formulate the decision problem as a multi-stage stochastic program \cite{bakker2020structuring}. This approach splits the decision problem into decision stages $n=1,\ldots, N$, where successive information disclosure is captured by a stochastic process: $\{\xi_1, \ldots, \xi_N \}$, and where each element $\xi_n$ of the process determines the parameters after the $n-$stage subproblem. The result of the program is a \textit{policy}, which describes what actions to take in sequential realizations of uncertain parameters.

Though redispatch is a continuous process, 
we use two simplifications to reduce the decision problem to a setting with two decision stages and two uncertainty stages (see Fig. \ref{fig: decision}). The first simplification is that the activation of CLC by the SO and the communication of prognosis by the CSP take place at the same time in decision stage 1. This is valid if the CSP does not gain significantly more insight into its expected EV load between 8:00 (CLC activation) and 14:30 or 15:30 (prognosis communication). We show in Section \ref{sec: EV fleet} that this is generally the case when forecasting EV fleet parameters. Second, we assume that the SO decides on only one moment to activate the RCs in the congested area, rather than doing so continuously. This is the second decision stage. This approach is close to the current practice of Dutch SOs (especially DSOs), as applying continuous redispatch in multiple congested areas is a very complex operational task.
The timing of that RC activation matters, as the available volumes and prices in the redispatch market vary over time, impacting both price and liquidity risk. Furthermore, later RC activation increases information gain between the two decision stages, potentially making RCs more attractive over CLCs. To include these effects, we parameterize the moment of RC activation by $\tau$, which can be on either day $D-1$ or day $D$ (see Fig. \ref{fig: decision}). If $\tau$ falls on day $D$ (e.g. 9:00 $D$), only CLCs and no RCs can be applied for the moments of delivery before $\tau$ (0:00-9:00 $D$). For these time steps, all the uncertainty is realized after the CLC decision stage. We now define $\tau$ more formally in the problem formulation.
\vspace{-10pt}

\subsection{Problem Formulation}
\label{sec: problem formulation}
Since only hourly redispatch products were observed in the GOPACS order book,  an hourly time resolution for the model ($\Delta t = 1 \mathrm{h}$) is selected. We denote the set of delivery hours on day $D$ as $\mathcal{D} = \{0, \ldots, 23\}$ and the set of decision stages of the problem as $\mathcal{N}=\{1, 2\}$. We now formally define the parameter $\tau \in \{-12, \ldots, 23\}$ as the moment of RC activation, the time stamp of the second decision stage of the problem. A value of $\tau = -12$ corresponds to RC activation at 12:00 on day $D\!-\!1$, whereas a value of $\tau = 12$ corresponds with RC activation at 12:00 on day $D$. It will be convenient to define $\mathcal{D}^-=\{d\in\mathcal{D}|d\leq \tau\}$ and $\mathcal{D}^+=\{d\in\mathcal{D}|d > \tau\}$ to distinguish moments of delivery on $D$ for which only CLC, or both CLC and RC activation is possible (see Fig. \ref{fig: decision}). The variables and parameters defined on $\mathcal{D}^-$ are only defined for one decision stage and one uncertainty stage, whereas the variables defined on $\mathcal{D}^+$ are defined for two decision stages and two uncertainty stages. 

\begin{figure}[!b]
\centering
\includegraphics[width=1.0\linewidth]{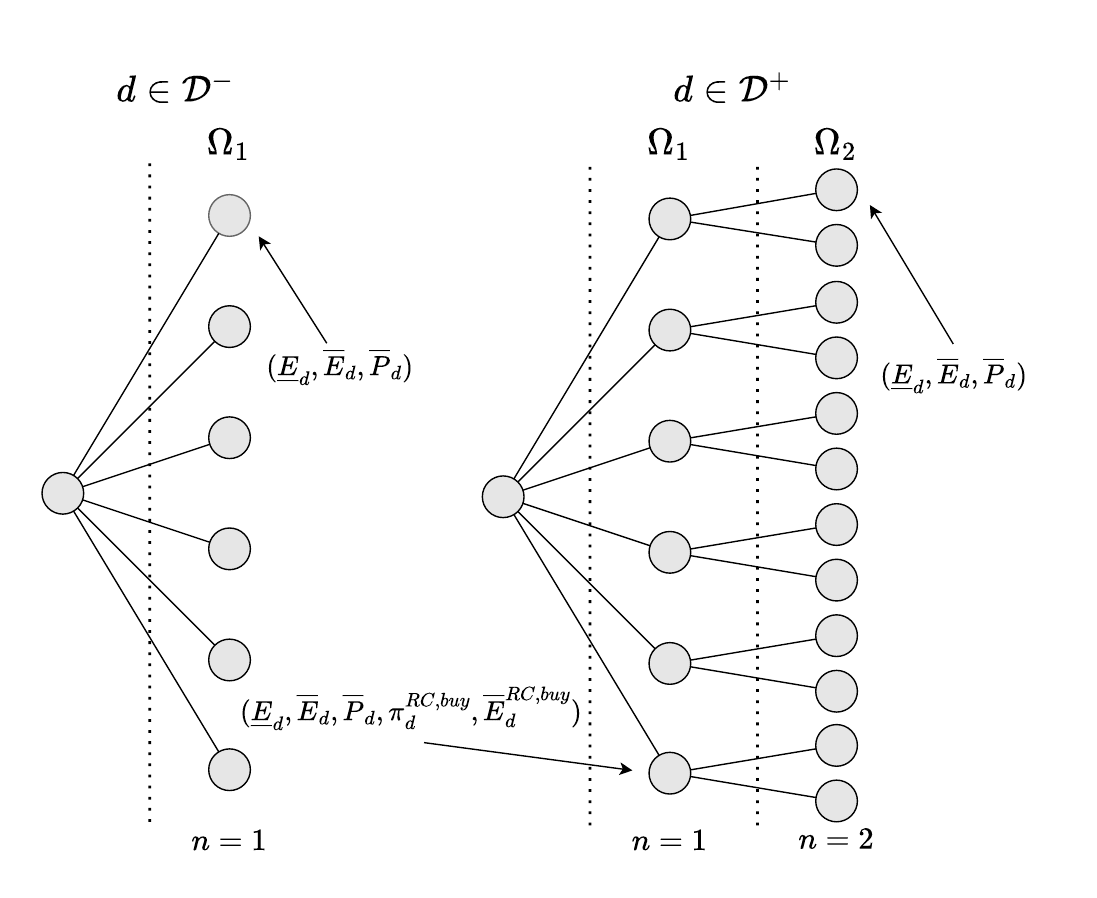}
\caption{Schematic representation of the scenario trees approximating the stochastic process $\{\xi_1, \xi_2\}$. For times of delivery $d\in \mathcal{D}^-$, the CLC is the only contract the SO can activate and all stochastic parameters realize after the CLC decisions stage. For times $d\in \mathcal{D}^+$, there are two decision and two uncertainty stages, resulting in two layers in the scenario tree. Every scenario in the scenario trees contains realizations for the EV parameters, whereas the redispatch parameters only enter scenario for the first stages for $d\in \mathcal{D}^+$. All symbols are explained in Section \ref{sec: problem formulation}}
\label{fig: scen tree}
\end{figure}

We define the amount of activated capacity limitation with the CLC as $\Delta P_d^{CLC}$ on $\mathcal{D}$ and the amount of downward redispatch procured from the RC as $\Delta E_d^{RC}$ on $\mathcal{D}^+$. We will not consider upward redispatch in this model, as our SO is only interested in preventing congestion caused by the load from EVs. 

We assume that the SO minimizes its congestion management costs, under capacity chance constraints:
  \addtolength\abovedisplayskip{-0.3\baselineskip}%
  \addtolength\belowdisplayskip{-0.3\baselineskip}%
\begin{equation}
  \begin{aligned}
    \min_{\Xi_1} \Biggl\{\sum_{d \in \mathcal{D}}\pi^{CLC} \Delta P^{CLC}_d +\min_{\Xi_2}\Biggl\{\E_{\xi_1}\left[\sum_{d\in \mathcal{D}^+} \pi_{d}^{RC} \Delta E_{d}^{RC}\right]\Biggr\}\Biggr\}, 
    \label{eq: OF}
  \end{aligned}
\end{equation}
\begin{align}
    \probP_{\xi_1}(P_d \geq L) & \leq 1 - \epsilon \label{eq: chance constraints a} && d \in \mathcal{D}^-, \\
        \probP_{\xi_2}(P_d \geq L) & \leq 1 - \epsilon \label{eq: chance constraints b} && d \in \mathcal{D}^+. 
\end{align}
Equation \eqref{eq: OF} shows that the SO minimizes its CLC costs and the expected RC costs. The RC activation is decided in the second stage, after $\xi_1$ has been realized. The chance constraints \eqref{eq: chance constraints a}-\eqref{eq: chance constraints b} limit the probability that the EV load $P_d$ exceeds the capacity limit $L$ of a critical asset to a value of $1-\epsilon$ for some small $\epsilon$. We thus assume that the value of $L$ is constant and known to the SO. 
Notice that we split the chance constraints based on the value of $\tau$ because the uncertainty for times of delivery $d\leq \tau$ and $d > \tau$ is realized when  $\xi_1$ and $\xi_2$ are disclosed, respectively (see Fig. \ref{fig: decision}).

Equations \eqref{eq: OF}-\eqref{eq: chance constraints b} are characterized by this stochastic process $\{\xi_1, \xi_2\}$. We model the process by the scenario trees presented in Fig. \ref{fig: scen tree}, with stages $n\in \mathcal{N}$ and scenarios
$\omega \in \Omega_n$ with probabilities $p_{\omega}$ at every stage. The objective function \eqref{eq: OF} can then be written in the \textit{extensive form} as:
\begin{equation}
  \begin{aligned}
     \min_{\Xi_1} \Biggl(&\sum_{d \in \mathcal{D}}\pi^{CLC} \Delta P^{CLC}_{d} +\\
    &\min_{\Xi_2}\Biggl(\sum_{\omega \in \Omega_1}\sum_{d\in \mathcal{D}^+} p_{\omega} \pi_{d, \omega}^{RC} \Delta E_{d, \omega}^{RC} \Biggr)\Biggr), 
    \label{eq: OF omega}
    \end{aligned}
\end{equation}
where the redispatch action $\Delta E_{d, \omega}^{RC}$ is now a recourse variable depending on the realization $\omega \in \Omega_1$ of the random variable $\xi_1$. The costs for the capacity reduction per MW $\pi^{CLC}$ are assumed to be constant and fixed in the contract, resulting in no price risk for the SO from the CLC. We similarly assume that the RC specifies a fixed price per MWh $\pi^{RC, sell}$ for which the CSP can sell its excess of energy in the redispatch market. As explained in Section \ref{sec: RC}, the resulting costs of redispatch for the SO are given by the spread between this (now deterministic) sell offer and the buy offer from some counter-party in GOPACS. This buy price is uncertain at the CLC decision stage, resulting in price risk for the SO from the RC. This results in the following price per MWh for redispatch for the SO:
\begin{align}
    \pi^{RC}_{d, \omega} = \max(0, \pi^{RC, sell} - \pi^{RC, buy}_{d, \omega})  && d \in \mathcal{D}^+, \omega \in \Omega_1.
\end{align}
The decision variables $\Delta P_{d}^{CLC}$ and $\Delta E_{d, \omega}^{RC}$ are bounded as follows:
\begin{align}
   0  \leq & \Delta P^{CLC}_{d} \leq \bar{P}^{CLC} - \ubar{P}^{CLC} &&  d\in \mathcal{D}, \label{eq: CLC bounds}\\
       z^{RC}_{d, \omega} \ubar{P}^{RC} \Delta t \leq & \Delta E^{RC}_{d, \omega} \leq z^{RC}_{d, \omega} \bar{E}^{RC, buy}_{d, \omega} &&  d\in \mathcal{D}^+, \omega \in \Omega_1,
    \label{eq: RC bounds}\\
    & z^{RC}_{d, \omega} \in \{0, 1\} &&  d\in \mathcal{D}^+, \omega \in \Omega_1,
\end{align}
where the value of $\bar{P}^{CLC}$ is the sum of the grid capacity of all the EV charging stations in the portfolio of the CSP, and  $\ubar{P}^{CLC} \leq L$ is some minimum capacity the CSP can be reduced to, as specified in the CLC. If redispatch is requested ($z^{RC}_{d, \omega} = 1$), the amount should be more than specified by the minimum bid size $\ubar{P}^{RC} = 0.1 \mathrm{MW}$ of the redispatch market, and less than the total buy volume in the redispatch market $\bar{E}^{RC, buy}_{d, \omega}$. This volume is considered a stochastic parameter unknown at the CLC decision stage, resulting in market liquidity risk for the SO.

We approximate the chance constraints \eqref{eq: chance constraints a}-\eqref{eq: chance constraints b} by bounding the Conditional Value at Risk (CVAR) of the EV load exceeding the limit $L$. The CVAR controls the expected magnitude of the limit violation, resulting in a restriction with respect to the original chance constraints \cite{nemirovski2007convex}. The advantage of this method is that it results in a set of linear constraints that control the magnitude of the capacity violation. The limitation is that it results in conservative solutions, which could result in the SO overprocuring flexibility for congestion management in theory.
Therefore, we will validate this approach for our problem in our out-of-sample analysis in Section \ref{sec: out-of-sample}. 

For the formulation of CVAR constraints, we distinguish between variables defined on $\mathcal{D}^-$ and $\mathcal{D}^+$ using $-$ and $+$ superscripts: 
\begin{align}
\zeta^-_{d, \omega} & \geq P^-_{d, \omega} - L - \eta_d &&  d \in \mathcal{D}^-, \omega \in \Omega_1,\\
\zeta^+_{d, \omega} & \geq P^+_{d, \omega} - L - \eta_d &&  d \in \mathcal{D}^+, \omega \in \Omega_2,\\
\zeta^-_{d, \omega} & \geq 0 &&  d \in \mathcal{D}^-, \omega \in \Omega_1,\\
\zeta^+_{d, \omega} & \geq 0 &&  d \in \mathcal{D}^+, \omega \in \Omega_2,\\
    \epsilon \eta_d &+ \sum_{\omega \in \Omega_1} p_{\omega}\zeta^-_{d, \omega} \leq 0 &&  d \in \mathcal{D}^-
    \label{eq: CVAR a} \\
 \epsilon \eta_d &+ \sum_{\omega \in \Omega_2} p_{\omega}\zeta^+_{d, \omega} \leq 0 &&  d \in \mathcal{D}^+ \label{eq: CVAR b}
\end{align}
where $\zeta^-_{d, \omega}$, $\zeta^+_{d, \omega}$, and $\eta_d$ are auxiliary variables. 

\begin{figure}[!t]
\centering
\includegraphics[width=\linewidth]{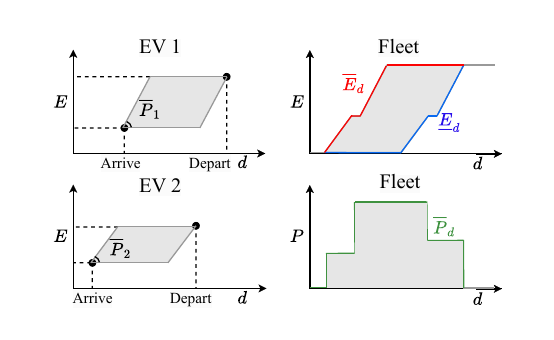}
\caption{Illustration of how the parameters $\bar{E}_d$, $\bar{E}_d$, and $\bar{P}_d$ can be calculated for a fleet of two cars from their respective charging session information. The two figures in the left column present the areas containing all possible charging schedules of the individual EVs. These areas are characterized by the maximum charge power of the respective EVs, and the arrival/departure times and state of charge. In the right column, the figures present the three fleet parameters based on the charging sessions in the left column in the energy-time and power-time plane.}
\label{fig: EVs}
\end{figure}

To estimate the effect of the contract activations, the SO needs state equations of the EV fleet and its general charging strategy. For the state equation, we use an aggregated virtual battery model from \cite{tang2016aggregated}. The model aggregates data of charging sessions from individual EVs (arrival/departure times and state of charge) into three fleet parameters that provide charging power and state of charge envelopes for the fleet. These parameters are: the cumulative \textit{fast-as-possible} energy charged $\ubar{E}_{d}$, the cumulative \textit{slow-as-possible} energy charged $\bar{E}_{d}$, and the instantaneous maximum charging power of the fleet $\bar{P}_{d}$. Fig. \ref{fig: EVs} demonstrates how these parameters can be calculated for a fleet consisting of two EVs, but for further details we refer the reader to \cite{tang2016aggregated}. In this work, these fleet parameters are considered to be uncertain at both the CLC and RC decision stage and are thus part of both $\xi_1$ and $\xi_2$. The modeling of the distributions of these stochastic time series is presented in Section \ref{sec: EV fleet}.


Let $E_{d, \omega}^-$ and $E_{d, \omega}^+$ denote the total charged energy of the EV fleet on $\mathcal{D}^-$ and $\mathcal{D}^+$, respectively. We can then write the following state equations for the fleet:
  \addtolength\abovedisplayskip{-0.2\baselineskip}%
  \addtolength\belowdisplayskip{-0.2\baselineskip}%
\begin{align}
   \ubar{E}^-_{d, \omega} \leq E^-_{d, \omega} &\leq \bar{E}^-_{d, \omega} &&  d\in \mathcal{D}^-, \omega\in \Omega_1, \label{eq: Emin a}\\
      \ubar{E}^+_{d, \omega} \leq E^+_{d, \omega} &\leq \bar{E}^+_{d, \omega} &&  d\in \mathcal{D}^+, n\in\mathcal{N}, \omega\in \Omega_n, \label{eq: Emin b}\\
          0 \leq P^-_{d, \omega} & \leq \bar{P}^-_{d, \omega}&&  d\in \mathcal{D}, \omega\in \Omega_1,\\
    0 \leq P^+_{d, \omega} & \leq \bar{P}^+_{d, \omega}&&  d\in \mathcal{D}, n\in\mathcal{N}, \omega\in \Omega_n,\\
   E_{d, \omega}^- = E_{d - 1, \omega}^- + & P_{d, \omega}^- \Delta t && d \in \mathcal{D}^- \setminus \{0\}, \omega\in \Omega_1 \\
E_{d, \omega}^+ = E_{d - 1, \omega}^+ + & E_{d, \omega}^+ \Delta t && d \in \mathcal{D}^+ \setminus \{0\}, n\in \mathcal{N}, \omega\in \Omega_n
\end{align}
In practice, the CSP controls the fleet and may have objectives that differ from those of the SO, as it can also participate in other electricity markets. However, the charging session data used in this study to derive the three EV parameters (see Section \ref{sec: EV fleet}) typically reflect fast-as-possible charging behavior. Therefore, we assume that the CSP adopts this strategy for the fleet when not providing congestion management. This means the CSP participates solely in congestion management and does not engage in other electricity markets, such as the day-ahead market. Given the formulation of the virtual battery model, the response function of the fleet after CLC activation and the realization of $\omega \in \Omega_1$ on e.g. $\mathcal{D}^-$ can then be expressed explicitly as: 
\begin{equation}
   P^-_{d, \omega} = \min{\left(\frac{1}{\Delta t}(\bar{E}^-_{d, \omega}-E^-_{d, \omega}), \bar{P}^-_{d, \omega}, \bar{P}^{CLC} - \Delta P_{d, \omega}^{CLC} \right)}, 
   \label{eq: charging}
\end{equation}
and similarly for $P_{d, \omega}^+$ on $d \in \mathcal{D}^+$. The charging power is simply the power necessary to get back to the state of charge under the fast-as-possible assumption ($\bar{E}^-_{d, \omega}$), unless that is impossible due to the maximum charging power of the fleet, or the imposed capacity limitation from the CLC. This non-convex relation can be reformulated in a mixed-integer linear form for all $d\in \mathcal{D}^-$ and $\omega \in \Omega_1$:
\begin{align}
    & P^-_{d, \omega} \leq \frac{1}{\Delta t}(\bar{E}^-_{d, \omega}-E^-_{d, \omega}) \\
    & P^-_{d, \omega} \leq \bar{P}^-_{d, \omega}\\
    & P^-_{d, \omega} \leq \bar{P}^{CLC} - \Delta P_{d}^{CLC}\\
    & P^-_{d, \omega} \geq \frac{1}{\Delta t}(\bar{E}^-_{d, \omega}-E^-_{d, \omega}) - M(1 - z^{-, a}_{d, \omega}) \\
    & P^-_{d, \omega} \geq \bar{P}^-_{d, \omega} - M(1 - z^{-, b}_{d, \omega})\\
    & P^-_{d, \omega} \geq \bar{P}^{CLC} - \Delta P_{d}^{CLC}- M(1 - z^{-, c}_{d, \omega})\\
    &z^{-, a}_{d, \omega} + z^{-, b}_{d, \omega} + z^{-, c}_{d, \omega} = 1\\
    &z^{-, a}_{d, \omega}, z^{-, b}_{d, \omega}, z^{-, c}_{d, \omega} \in \{0, 1\},
\end{align}
where the value of $M$ needs to be larger than all possible values of $P^-_{d, \omega}$, and where the superscripts $a, b, c$ refer to the three cases in the min operator in equation \eqref{eq: charging}. The same relations hold for $P^+_{d, \omega}$. Since the prognosis of the CSP is determined directly after CLC activation in our model, the values of $P_{d, \omega}^-$ and $P^+_{d, \omega}$ together form the prognosis the CSP communicates to the SO for $d\in\mathcal{D}^-$ and $d \in \mathcal{D}^+$, respectively.

To calculate the charging powers after the RC activation for $d > \tau$ (that is, $P^+_{d, \omega}$ for $\omega \in \Omega_2$), the same relations can be used, but a fourth case for redispatch should be added. This results in the following set of constraints for all $d\in \mathcal{D}^+$ and $\omega \in \Omega_2$:
\begin{align}
    & P^+_{d, \omega} \leq \frac{1}{\Delta t}(\bar{E}^+_{d, \omega}-E^+_{d, \omega})\\
    & P^+_{d, \omega} \leq \bar{P}^+_{d, \omega}\\
    & P^+_{d, \omega} \leq \bar{P}^{CLC} - \Delta P_{d}^{CLC}\\
    & P^+_{d, \omega} \leq P^+_{d, \omega_1} - \frac{1}{\Delta t} \Delta E^{RC}_{d, \omega_1} \label{eq: redispath 1}\\
    & P^+_{d, \omega} \geq \frac{1}{\Delta t}(\bar{E}^+_{d, \omega}-E^+_{d, \omega}) - M(1 - z^{+, a}_{d, \omega})\\
    & P^+_{d, \omega} \geq \bar{P}^+_{d, \omega} - M(1 - z^{+, b}_{d, \omega})\\
    & P^+_{d, \omega} \geq \bar{P}^{CLC} - \Delta P_{d}^{CLC}- M(1 - z^{+, c}_{d, \omega})\\
    & P^+_{d, \omega} \geq P^+_{d, \omega_1} - \frac{1}{\Delta t} \Delta E^{RC}_{d, \omega_1}- M(1 - z^{+, d}_{d, \omega})\label{eq: redispath 2}\\
    &z^{+, a}_{d, \omega} + z^{+, b}_{d, \omega} + z^{+, c}_{d, \omega} + z^{+, d}_{d, \omega} = 1\\
    &z^{+, a}_{d, \omega}, z^{+, b}_{d, \omega}, z^{+, c}_{d, \omega}, z^{+, d}_{d, \omega} \in \{0, 1\} \label{eq: last equation},
\end{align}
where scenario $\omega_1 \in \Omega_1$ is the predecessor of $\omega \in \Omega_2$ in the scenario tree. The new equations \eqref{eq: redispath 1} and \eqref{eq: redispath 2} limit the charging power to the prognosis minus the activated redispatch and couple the charged power after the two decision stages. The equations \eqref{eq: OF omega}-\eqref{eq: last equation} present the complete decision model, being a mixed-integer linear program (MILP).
\subsection{Solution Strategy}
The MILP described in equations \eqref{eq: OF omega}-\eqref{eq: last equation} can be challenging to solve for an increasing number of scenarios. 
In principle, decomposition techniques 
can be applied to multi-stage problems with coupling constraints like CVAR constraints \eqref{eq: CVAR a}-\eqref{eq: CVAR b}. However, the results in this paper are obtained by solving the decision model on $\sim 10^3$ distinct scenario trees, and achieving consistent convergence with these algorithms proved to be challenging. We therefore solve the extensive form of the problem using Gurobi \cite{gurobi} with problem-specific heuristics. For example, if none of the scenarios resulted in congestion before 16:00 $D$ (as was often the case), no congestion management was necessary for $d < 16$, and all those variables could be precalculated. This reduced the size of the problem two-thirds. We also provided a warm start solution calculated for a similar problem with a zero tolerance for congestion ($\epsilon = 0$). This problem is significantly easier to solve than the original problem. The resulting solving time was at most 8 hours per scenario tree on an Intel® Xeon® CPU ES-2687W v4 with 24 cores and 128 GB of RAM with a MIP gap of $0.1\%$.  
\section{Uncertainty Modeling}
\label{sec: scenario generation}
This section outlines the scenario generation approaches for the EV fleet and redispatch market parameters, which are modeled independently and discussed in separate subsections. Afterwards, the structure of the resulting scenario trees is discussed. 

\subsection{EV Fleet}
\label{sec: EV fleet}
Section \ref{sec: problem formulation} described how stochastic time series $\ubar{E}_d$, $\bar{E}_d$, and $\bar{P}_d$ parameterize the EV fleet for the chosen virtual battery model and that these parameters are uncertain at both the CLC and RC decision stage. This means that the SO should create scenarios for these parameters at both decision moments, taking place at 8:00 $D-1$ and at $\tau$, respectively. Since we analyze the model results for various values of $\tau$, we effectively need to model how the probability distributions of $\ubar{E}_d$, $\bar{E}_d$, and $\bar{P}_d$ change while approaching real-time delivery. To do so, we base our approach on the data set of domestic charging in the UK from 2017 \cite{evdata}, containing more than three million charging sessions from 25.125 domestic charging points. This data set was selected because it is open; it contains enough charging points to consider large fleets; and because domestic charging in the UK has similar dynamics as in the Netherlands in terms of arrival/departure statistics and aggregated load curves \cite{sadeghianpourhamami2018quantitive}\cite{quiros2015statistical}.

\begin{figure}[!t]
\centering
\includegraphics[width=0.95\linewidth]{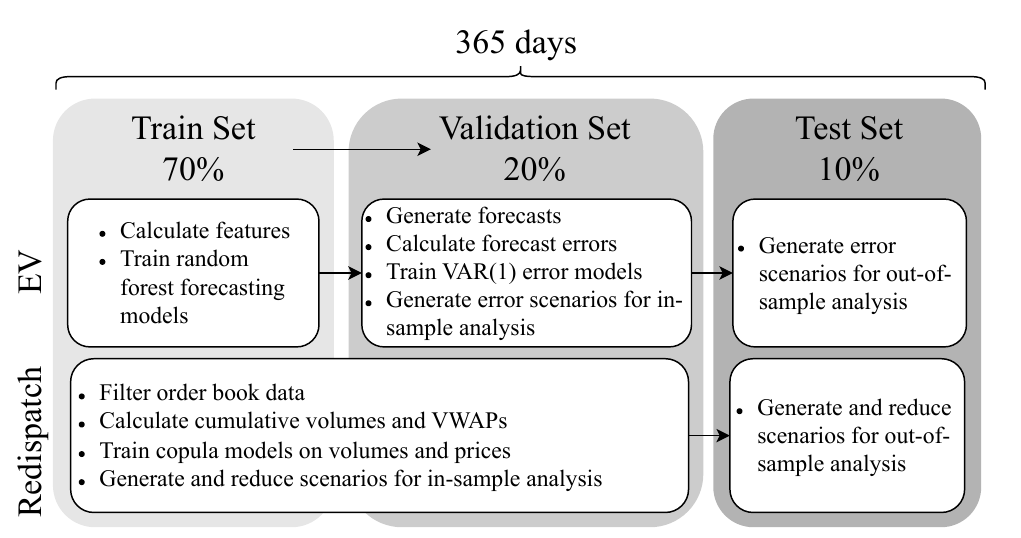}
\caption{General overview of the scenario generation approaches for both the EV fleet and redispatch market parameters.The 2017 EV charging session data and the 2023-2024 redispatch market order book data were aligned by corresponding calendar days.}
\label{fig: SG}
\end{figure}

The scenario generation approach is illustrated in Fig.~\ref{fig: SG} and consists of three steps: 1) train a machine learning model to predict $\ubar{E}_d$, $\bar{E}_d$, and $\bar{P}_d$ for various \textit{forecast times} and fleet sizes on 70\% of the charging session data (the training set); 2) evaluate the forecast errors on 20\% of the data (the validation set) to analyze forecasting errors and train error models on them; and 3) generate scenarios (forecasts + errors) for the remaining 10\% of the days (the test set) for the out-of-sample analyis.

Training, validation, and test sets are randomly selected from the 365 days in the EV data set.
The forecast time refers to the moment the scenarios are created (i.e. 8:00 $D-1$ and all possible values of $\tau$). For the fleet size $V$, the values $2.500, 10.000$, and $25.000$ are considered. The fleet size is important because it determines the scale of the congestion problem and how it relates to available counter volumes in the redispatch market. Parameters for larger fleets might also be easier to forecast because of the higher level of aggregation. The forecast time is important because the improvement of forecasts for later $\tau$ reflects the advantage of solving congestion intraday with RCs over using CLCs early.

\begin{figure}[!t]
\centering
\includegraphics[width=\linewidth]{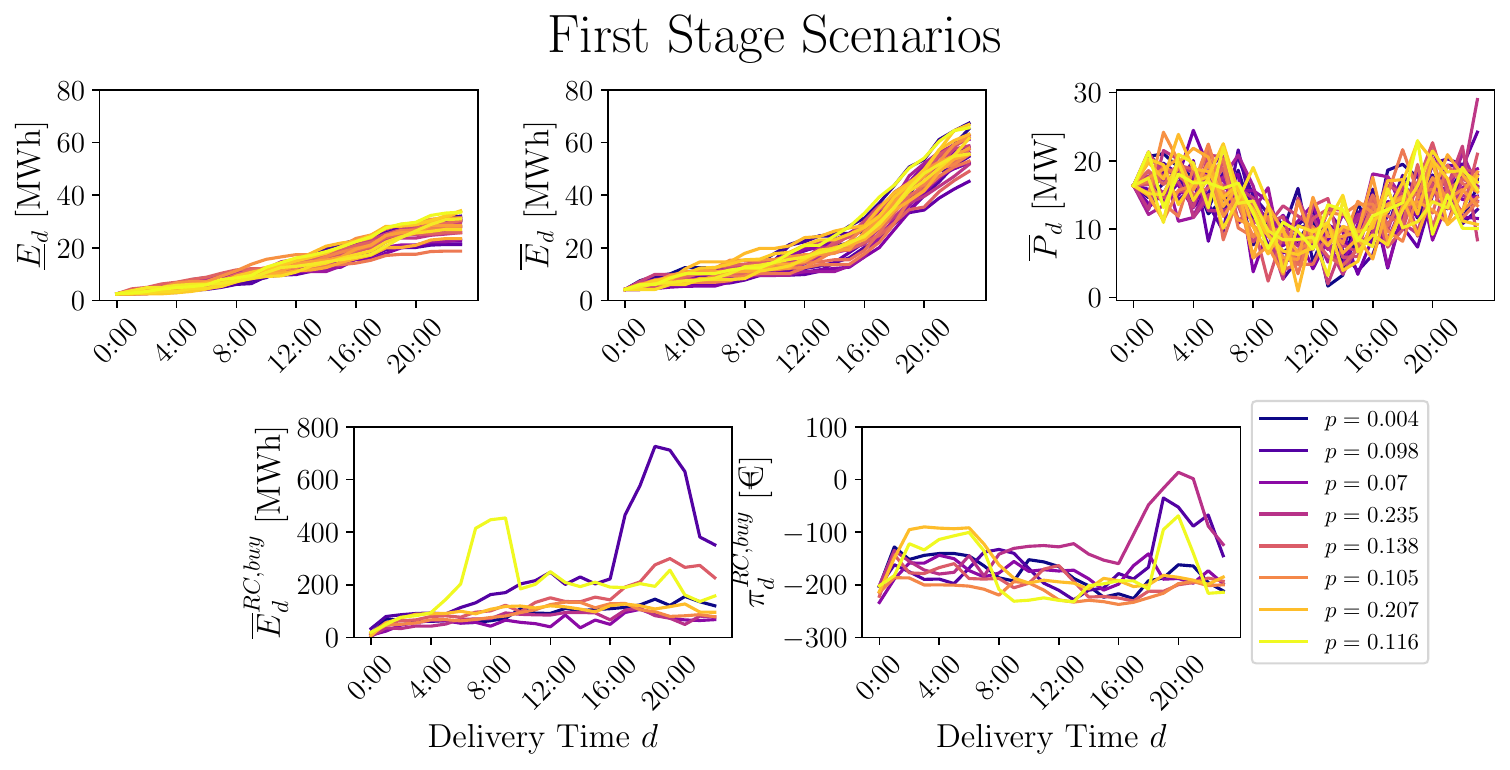}
\caption{Generated scenarios for the first decision stage for the 9th of October for a fleet size of $V=10.000$ and a redispatch contract activation at 23:00 $D\!-\!1$. For the three EV parameters in the first row, 28 equiprobable scenarios are presented. In the second row, eight scenarios and probabilities are presented for the buy volumes and prices in the redispatch market.}
\label{fig: scenario plot}
\end{figure}

For the training step, all daily values for $\ubar{E}_d$, $\bar{E}_d$, and $\bar{P}_d$ were calculated for the days in the training set. In a paper from 2024, Brinkel \textit{et al.} \cite{brinkel2023novel} investigated several machine learning methods to forecast these parameters for the virtual battery model adopted in this paper. We used their best-performing random forest (RF) model and evaluated the forecast errors for the various forecast times on the 73 days in the validation set. The results showed that the mean relative forecast errors did not exceed $0.2$ and decreased significantly for forecast times after 0:00 $D$ (data not shown). This indicates that using RCs intraday over day-ahead CLCs can be attractive. Furthermore, no clear decrease of relative forecasting errors was observed for increasing fleet sizes. The forecast errors were generally normally distributed with a mean close to $0$. In addition, auto- and cross-correlation plots for the three error time series show only a significant correlation at lag 1. For these reasons, a set of VAR($k=1$) models was trained on the errors in the validation set. There, the Bayesian Information Criterium (BIC) reaffirmed that $k=1$ was the most appropriate for $85.4\%$ of the error models. Having established the forecast and error models, scenario sets of size $|\Omega_{EV}|$ were created for all days in the validation set. In the in-sample stability analysis, the appropriate number of $|\Omega_{EV}|$ is determined (see Section \ref{sec: in-sample}). This number of first and second stage scenarios is created for the days in the test set for the out-of-sample analysis.  Fig. \ref{fig: scenario plot} presents an example of generated scenarios for $|\Omega_{EV}| = 28$, for a forecast time of 23:00 $D\!-\!1$, and a fleet consisting of $V=10.000$ vehicles.

\subsection{Redispatch Market}
\label{sec: redispatch market}
The redispatch buy volumes $\bar{E}^{RC, buy}_d$ and corresponding prices $\pi^{RC, buy}_d$ model the redispatch market for this decision problem. These parameters are considered to be uncertain at CLC activation stage only. The redispatch volumes and prices realize after the first stage and are available for RC activation. We repeat that RC activation is only possible for delivery times $d \in \mathcal{D}^+$, so the parameters only enter the scenario trees for those $d$ (see Fig. \ref{fig: scen tree}). 

To estimate the distributions of these parameters, ETPA, one of the two nominated electricity market operators providing access to GOPACS, shared the limit order book for the redispatch products. The data set contains all information about the orders submitted between May 2023 to June 2024 and was anonymized such that all information about the market parties submitting the orders or their location in the grid was removed. More than 96\% of the total cleared volumes presented on the GOPACS website \cite{GOPACS} could be traced back to order book from ETPA, indicating that their order book provides a good overview of the total market.

At $\tau$, the SO activates the RC and the CSP submits its sell order to GOPACS. This limit order can be (partially) matched with any buy order active in the order book between $\tau$ and the GC of the product, as long as it is submitted from outside the congested area. In a typical market, a price spread limits matching, but in this redispatch market the SO can cover the price spread if matching is necessary. The timing of RC activation determines the trading window and influences the available volumes and prices. We thus need to quantify the buy prices and volumes for the various trading windows starting at $\tau$. Although we do not know the location of the orders, we consider all orders in the data set to be submitted from outside the simulated congested area. The largest fleet considered in this study represents only $5.4\%$ of the national fleet size \cite{elaadnl}. Assuming a similar spatial distribution between redispatch counter-volume and EVs, the overestimation of counter-volume from outside the congested zone is about 5\% too. 

We consider both matched and unmatched orders for volume quantification, but we apply two filters to prevent volume overestimation. First, to reduce including orders related to order book manipulation, we remove orders existing for shorter than 15 minutes in the order book. Second, if orders with the same price and volume are resubmitted after retraction or expiration, we only count the first order. This prevents double-counting the same volumes. The total volume of the resulting set of buy orders is taken to be the maximum counter volume $\bar{E}^{RC, buy}_d$ for our CSP. The volume-weighted average price (VWAP) of these orders is taken to be the value of $\pi^{RC, buy}_d$. Applying the VWAP is a standard method for aggregating buy and sell volumes at different price levels into a single representative price. However, this approach may overestimate prices when only a small amount of redispatch volume is involved. More advanced methods introduce multiple price levels corresponding to different volume quantiles \cite{shinde2022multistage}, but this comes at the cost of a linear increase in the scenario tree size with the number of price levels.

\begin{figure}[!b]
\centering
\includegraphics[width=0.9\linewidth]{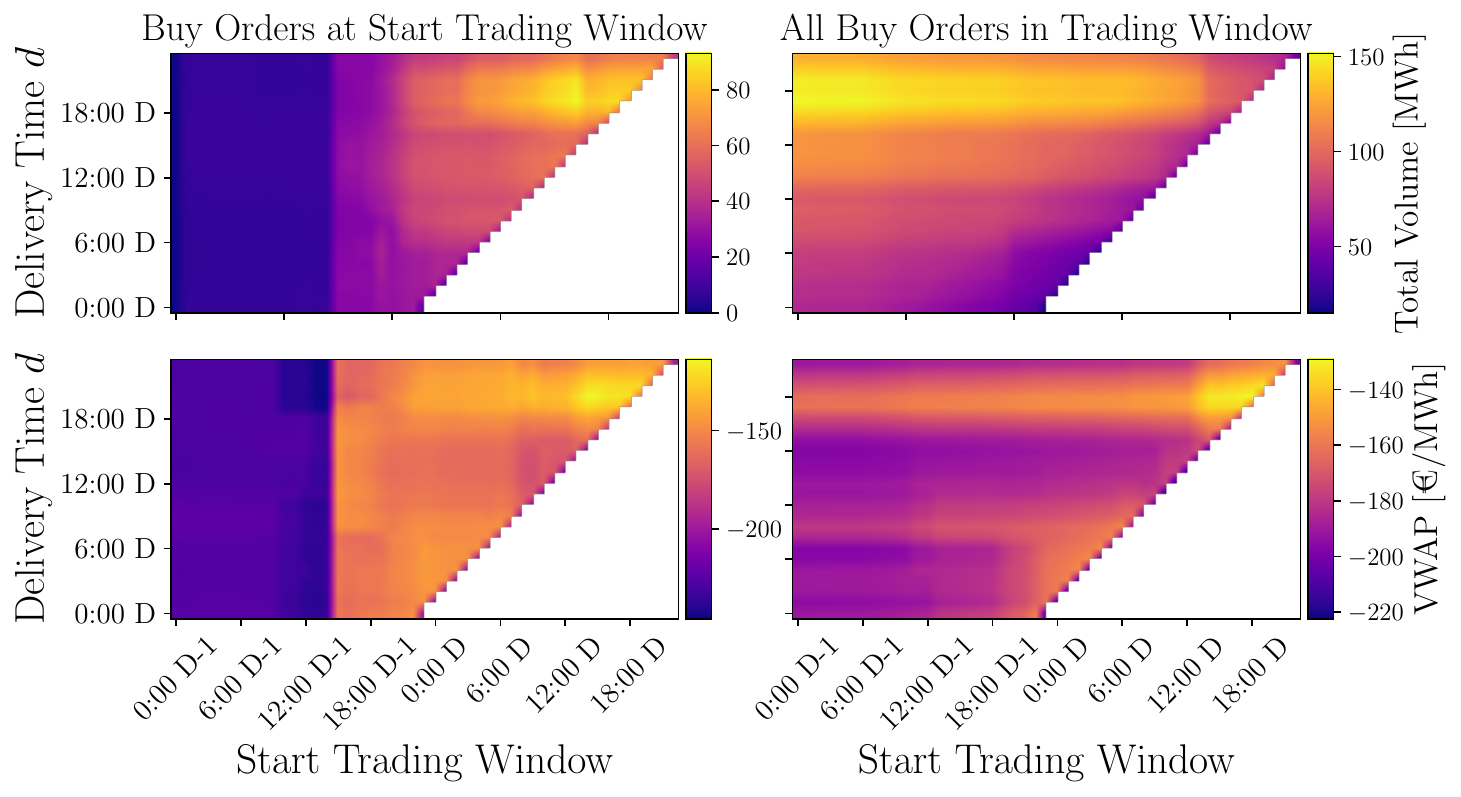}
\caption{The mean total volumes and volume-weighted average prices over the trading windows for redispatch products in the train/validation set. The column on the right presents the parameters over all orders in the trading window, whereas the column on the left includes only the subset of orders active at the very start of the window.}
\label{fig: order book}
\end{figure}

Fig. \ref{fig: order book} presents the mean of these volumes and prices for various trading windows. The means are calculated over all the days in the train and validation set (see Fig. \ref{fig: SG}). The column on the right presents the parameters over all orders in the trading window, whereas the column on the left includes only the subset of orders active at the very start of the window. We observe in the top-left plot that there is almost no buy volume in the order book before 15:00 $D\!-\!1$, being the gate-opening time of the intra-day market. This shows that though there are separate order books for products traded on the intraday and redispatch market, market parties typically do not start bidding until both markets are open. We also observe in the same figure that products for 18:00 $D$-22:00 $D$ are traded the most, and typically in the hours leading up to delivery. In the bottom-left plot, we see that these orders also have the highest buy prices, but that they are negative on average still. The negative prices suggest that the market is still very immature and that market parties take advantage of the SOs paying the spread between buy and sell orders. This then results in high costs for the SO per unit of redispatch. The top-right and bottom-right figures show the statistics for $\bar{E}^{RC, buy}_d$ and $\pi^{RC, buy}_d$ respectively. The first shows for every product how the total potential counter volume decreases as activating RCs is delayed (i.e. shorter trading windows). Delaying RC activation thus increases the risk of not finding enough counter-volume.

\begin{figure}[!t]
\centering
\includegraphics[width=0.8\linewidth]{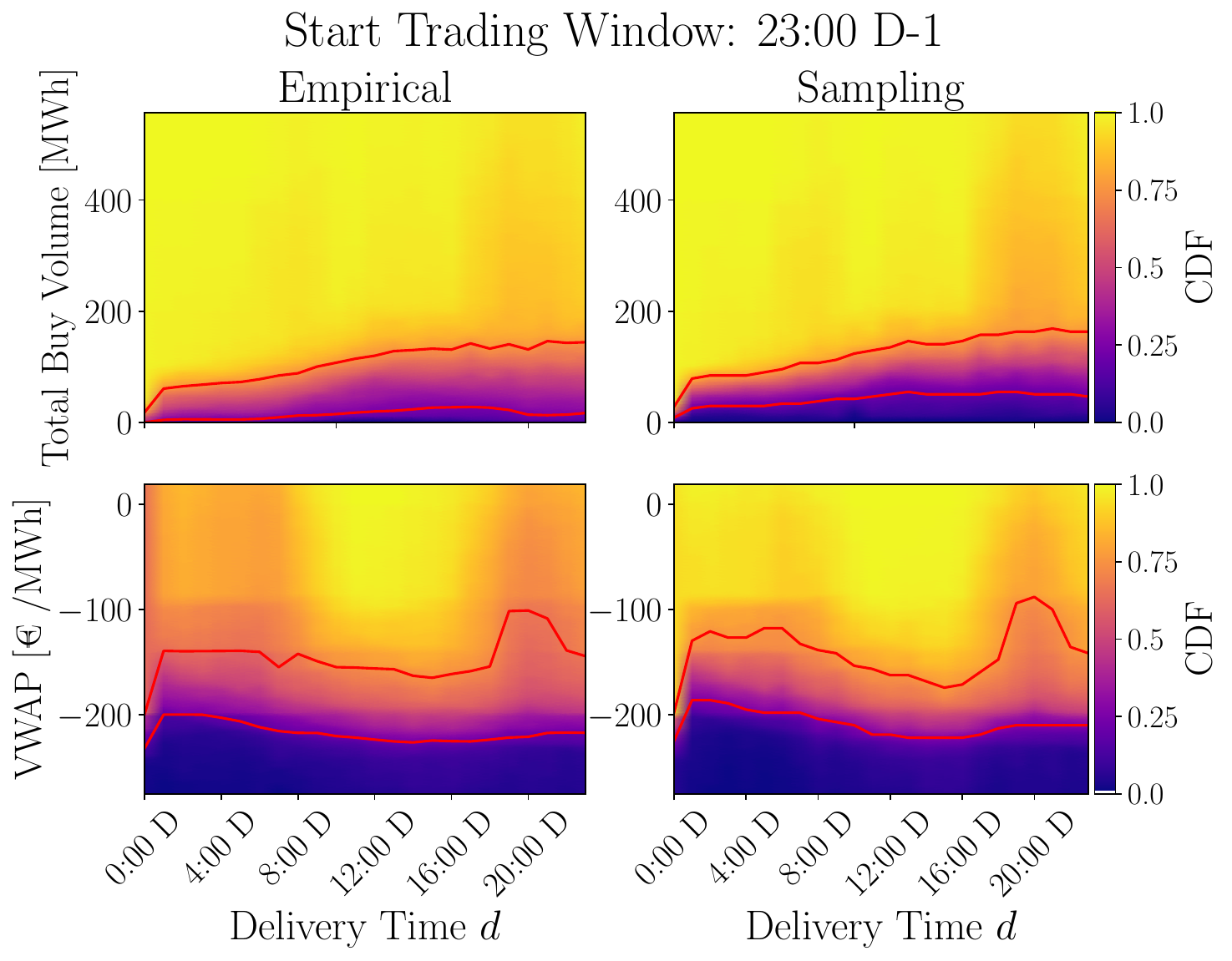}
\caption{The Cumulative Distribution Function (CDFs) for buy volumes and prices for all trading windows in the training and validation set, starting from 23:00 $D\!-\!1$. The first column shows the empirical CDFs from the data set, whereas the second column presents the CDFs obtained after sampling 10.000 samples from the trained gaussian or student-t copulas. Two red lines show the 25\% and 75\% quantiles.}
\label{fig: CDF heatmap}
\end{figure}

Fig. \ref{fig: CDF heatmap} presents the empirical cumulative distribution functions (CDFs) for the volumes and prices for all products traded in trading windows starting at 23:00 $D\!-\!1$ in the left column. To model this $2\times 24$ dimensional probability distribution, we use either Gaussian or student-t copulas, depending on the best fit using the BIC metric. For 37 out of the 47 trading windows considered, the student-t copula was selected. The right column of Fig. \ref{fig: CDF heatmap} shows the CDFs obtained from sampling 10.000 samples from the trained student-t copula. The 25\% and 75\% quantiles for all products are shown in red to illustrate that the copula seems to capture the general characteristics of the empirical distributions well. We also observe that we overestimate the buy volume of the lowest 25\% of the instances. This tells us that using this copula can result in an underestimation of the liquidity risk of RCs, potentially resulting in residual congestion for the SO in practice. 

To generate representative samples, the 10.000 samples were reduced to $|\Omega_{RC}|$ scenarios using the fast-forward reduction technique \cite{heitsch2003scenario}. In our in-sample stability analysis in Section \ref{sec: in-sample}, we will determine the appropriate value of $|\Omega_{RC}|$ for our problem. Fig. \ref{fig: scenario plot} illustrates the scenarios for our example for $|\Omega_{RC}|=8$.

\subsection{Scenario trees}
\label{sec: scenario trees}
Given the fleet size $V$ and the redispatch timing $\tau$, we can now construct the scenario trees presented in Fig. \ref{fig: scen tree}. For times of delivery $d \in \mathcal{D}^-$, only CLC activation is possible and the redispatch market is not important. That decision stage takes place at 8:00 $D-1$ and the EV parameters are uncertain. A number of $|\Omega_{EV}|$ scenarios for $\ubar{E}_d, \bar{E}_d, \bar{P}_d$ with a forecast time of 8:00 $D-1$ are generated, using the methods from Section \ref{sec: EV fleet}. For times of delivery times $d \in \mathcal{D}^+$, both CLC and RC activation is possible, resulting in two decision stages. At the CLC decision stage, the redispatch market parameters are relevant and uncertain. On top of the $|\Omega_{EV}|$ EV scenarios,  $|\Omega_{RC}|$ scenarios are generated for $\bar{E}^{RC, buy}_d$ and $\pi^{RC, buy}_d$ for a trading window starting at $\tau$, using the methods from Section \ref{sec: redispatch market}. Since the two processes are considered independent, this results in $|\Omega_{EV}| \times |\Omega_{RC}|$ scenarios in the first stage. For the RC activation, taking place at $\tau$, the redispatch parameters have realized and only the EV parameters are uncertain. A number of $|\Omega_{EV}|$ scenarios are generated for the EV parameters with a forecast time of $\tau$.
\section{Results}
\label{sec: results}
\begin{figure}[!t]
\centering
\includegraphics[width=0.6\linewidth]{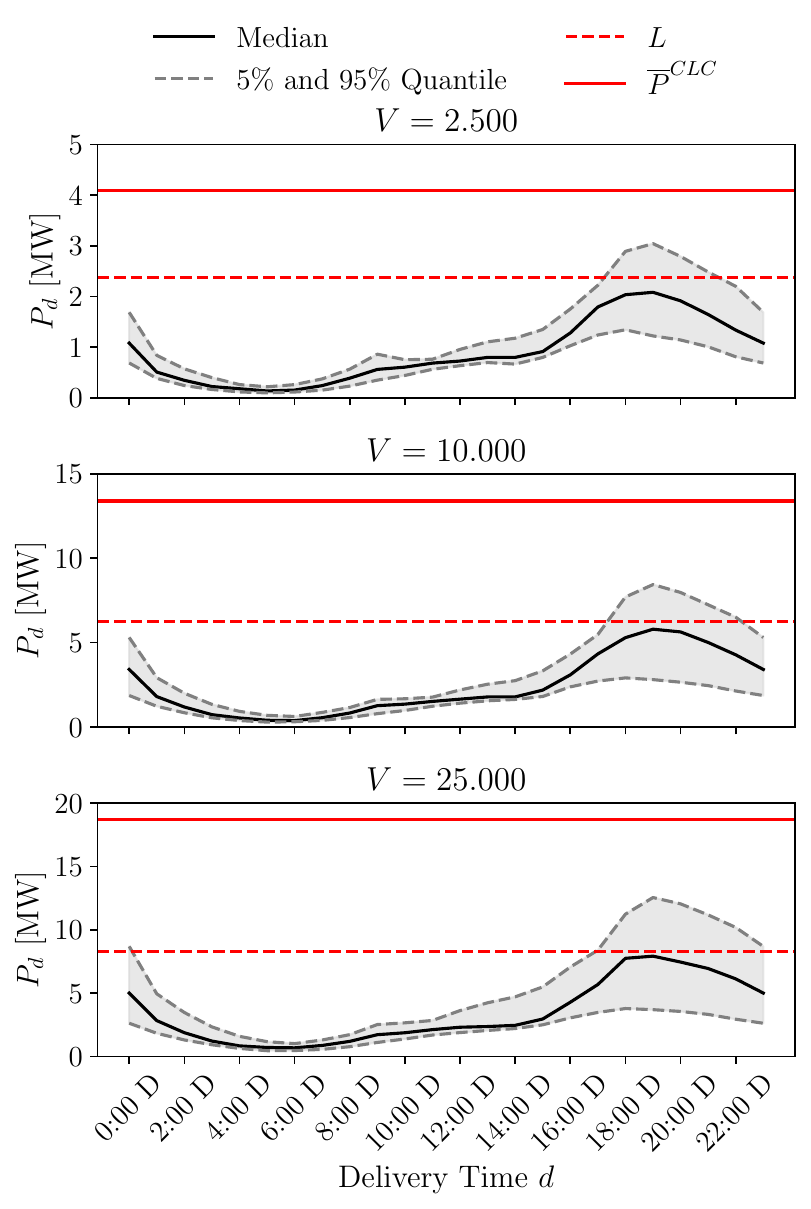}
\caption{The baseline charging power according to the \textit{fast-as-possible} charging strategy for the three different fleet sizes. The median values and the 5\%-95\% quantile range are presented, together with values from Tab. \ref{tab: dec params}.}
\label{fig: capacities}
\end{figure}

Tab. \ref{tab: dec params} presents values for the parameters in the model. The CSP's compensation for the CLC and the RC is based on the average DA price in the Netherlands in 2023. The minimum bid size of the redispatch market is the current value of $0.1$ MW and the congestion probability is 0.05. The total connection capacity of the EVs $\bar{P}^{CLC}$ and the available capacity for the EV fleet on the critical asset $L$ scale with the fleet size $V$. To ensure feasibility, the values of $L$ and $\bar{P}^{CLC}$ are selected to be the maximum charging power under the \textit{slow-as-possible} strategy over all predictions and scenarios, respectively. Fig. \ref{fig: capacities} shows these values compared to the statistics of the desired charging power. The value of $\ubar{P}^{CLC}$ is set equal to $L$ as setting lower values is not beneficial to the CSP or the SO.  

\begin{table}[h!]
  \centering
  \caption{Table of Decision Model Parameters}
  \label{tab: dec params}
  \begin{tabular}{l l|l|l|l}
    \hline  
    $\pi^{CLC}$ & [€/MW]& 100.00  & $\pi^{RC, sell}$ [€/MWh]& 100.00\\ 
    $\ubar{P}^{RC}$ & [MW]& 0.1 & $\epsilon$ \hspace{25pt} [-] & 0.05 \\ 
    \hline
    & & $V=2.500$ & $V=10.000$ & $V=25.000$ \\ 
    \hline
    $L$ & [MWh]& 2368.64& 6249.68 & 8294.26\\ 
    $\ubar{P}^{CLC}$ & [MWh]& 2368.64& 6249.68 &  8294.26\\ 
    $\bar{P}^{CLC}$ & [MWh]& 4098.80 & 13385.67 & 18687.31\\ 
    \hline
  \end{tabular}
\end{table}

\subsection{In-sample Stability Analysis}
\label{sec: in-sample}
\begin{figure}[!t]
\centering
\includegraphics[width=\linewidth]{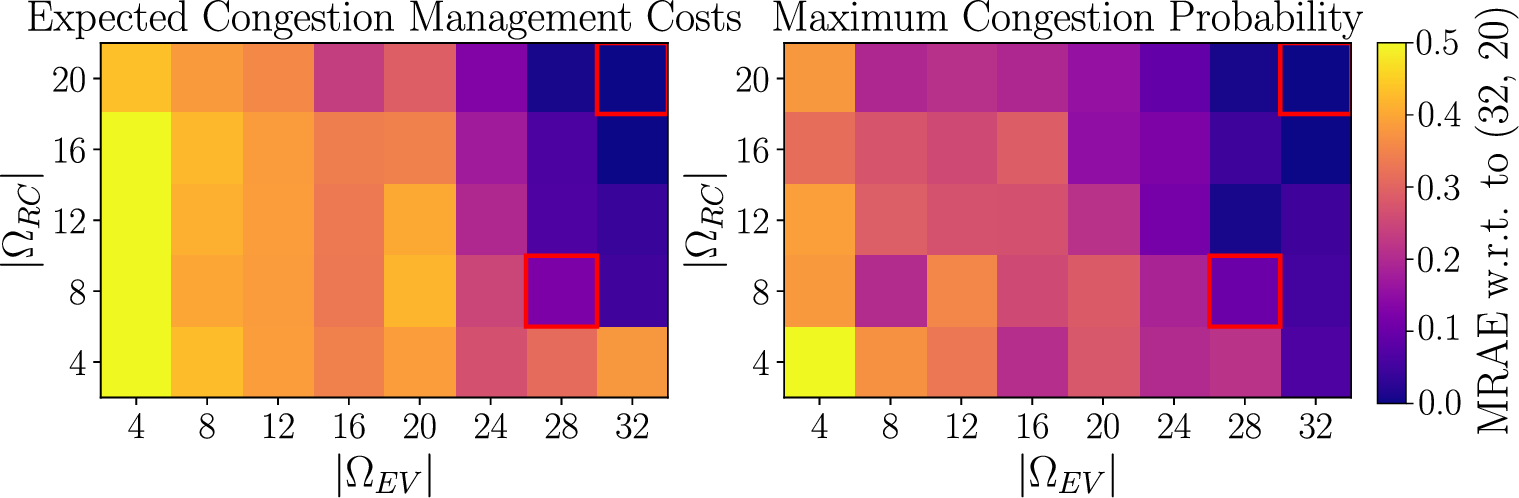}
\caption{The Mean Relative Absolute Error (MRAE) for the expected congestion management costs and the maximum congestion probability over day $D$.  The means are taken over the 73 days in the validation set. The reference for the error is the solution of the large problem including $|\Omega_{EV}| = 32$ and $|\Omega_{RC}|  = 20$ scenarios. Its MRAE, together with the selected values of $|\Omega_{EV}| = 28$ and $|\Omega_{RC}|  = 8$ are marked in red.}
\label{fig: insample}
\end{figure}

The objective of this in-sample stability analysis is to determine the appropriate sizing of the scenario tree presented in Fig. \ref{fig: scen tree}. This analysis can be performed for multiple values of $V$ and $\tau$, but we restrict ourselves to the setup used as an example throughout this article: $V=10.000$ and redispatch activation at 23:00 $D-1$. For all days in the validation set, we calculate the maximum congestion probability over $d$ and the expected congestion management costs for scenario trees with various values of $|\Omega_{EV}|$ and $|\Omega_{RC}|$. This implies solving the decision model on 2.880 scenario trees. Fig. \ref{fig: insample} presents the mean relative absolute errors of these two metrics compared to a big reference problem with $|\Omega_{EV}| = 32$ and $|\Omega_{RC}|  = 20$ scenarios. The reference problem should give the most accurate estimates of the metrics.

Fig. \ref{fig: insample} presents the resulting Mean Relative Absolute Error (MRAE) values. We observe that generally the MRAEs decrease for both metrics for larger scenario sets. Moreover, we observe that for $|\Omega_{EV}| = 28$ and $|\Omega_{RC}| = 8$ (marked in red) small MRAE values of 0.124 and 0.068 can be obtained for the expected congestion management costs and the maximum congestion probability, respectively. For the out-of-sample analysis, we use these numbers of scenarios, reducing the scenario tree by nearly $70\%$ compared to the large reference problem.

\subsection{Out-of-sample Analysis}
\label{sec: out-of-sample}
\begin{figure}[!t]
\centering
\includegraphics[width=1.0\linewidth]{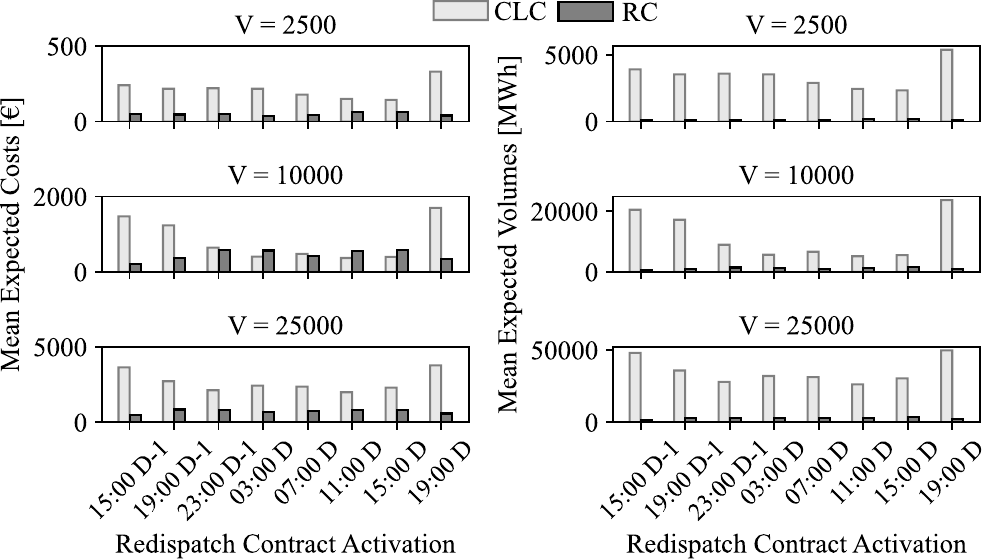}
\caption{The mean expected costs and volumes for capacity limitation contract (CLC) and redispatch contract (RC) activation for various RC activation times and fleet sizes $V$. The activated volumes for CLCs refer to the capacity limitation in terms of energy. The means are calculated over the 36 days in the test set.}
\label{fig: costs volume}
\end{figure}
\subsubsection{Exploring Policy Sensitivity to Model Parameters}
We start the out-of-sample analysis by studying the calculated policy. Fig. \ref{fig: costs volume} presents the mean expected costs and activated volumes for the CLC and RC activation for various fleet sizes and RC activation moments. The means are calculated over the 36 days in the test set. The first observation is that the expected optimal ratio for CLCs versus RC is typically significantly smaller in terms of costs than in terms of volumes. This is because redispatch is more expensive per unit, as the SO needs to pay the spread between $\pi^{RC, sell}$ and the counter buy order that is negative on average.
However, the figures show that this does not imply that CLCs should be the sole instrument to solve congestion with. Among the results of the individual days, two main reasons were observed. First, CLCs can be relatively expensive if the expected load is just above $L$. Invoking a small amount of redispatch is then still cheaper than paying $\approx \pi^{CLC}(\bar{P}^{CLC} - L)$. Second, if the prognosis of the CSP was relatively low compared to the latest congestion forecast, the CSP can be curtailed very cheaply by requesting the minimum possible amount of re-dispatch $\ubar{P}^{RC}\Delta t$ with respect to the low prognosis. This forces the CSP to consume little under the old and inaccurate prognosis, without receiving much compensation. This effect is known as the "0 kW paradox" \cite{ziras2021baselines}. Note that this effect can be partly mitigated by the CSP updating the prognosis over time, which is obligated according to the Dutch grid codes under some conditions, but not taken into account in this model.

If we look at the effect of the fleet size in Fig. \ref{fig: costs volume}, we observe that for the small fleet size $V = 2.500$ the CLC is preferred over the RC. The main reason is that Fig. \ref{fig: capacities} shows that the expected congestion is of similar size as the minimum redispatch bid size of $\ubar{P}^{RC}=0.1$MW. Being forced to buy redispatch for this minimum amount for small fleets can result in overprocuring flexibility and overpaying. This minimum bid size thus reduces the effectiveness of RCs in the lower grid levels. For $V=10.000$, this problem is smaller, resulting in a greater share of RC activation. Furthermore, for this fleet size, the required volumes were always expected to be available in the redispatch market. This was not always the case for the largest fleet of $V=25.000$ cars, resulting in lower RC activation. In several of the test days, congestion was expected with high probability, together with a non-zero probability of very low market liquidity. The small accepted congestion probability of $\epsilon = 0.05$, in combination with the conservative CVAR approximation, resulted in not activating the RC.

If we look at the effect of the RC activation moment on the costs in Fig. \ref{fig: costs volume}, we see that the congestion management costs generally decrease for later RC activation. This is as expected, as the RC activation is then informed by better forecasts and less congestion was oversolved. This makes RCs more attractive. We see the opposite effect if the SO has perfect information of the future (an oracle). Tab. \ref{tab: oracle} presents for $V=25.000$ and redispatch at 11:00 $D$ how the mean costs for the instruments compare for our policy compared with the oracle policy. We observe that the oracle policy is significantly cheaper and that it uses significantly less of the RC. The more expensive RC is now not necessary to hedge the risk of having a bad forecast at the CLC decision stage. This shows that the presence of uncertainty comes with a cost and impacts the optimal distribution of resources over the instruments. 

\begin{table}[h!]
  \centering
  \caption{Average Congestion Management Costs for $V=25.000$ and redispatch activation at 11:00 $D$}
  \label{tab: oracle}
  \begin{tabular}{l|l|l}
     & Stochastic Policy & Oracle Policy\\ 
    \hline  
    CLC Costs [€] & 1992.44 & 1055.23\\ 
    RC Costs [€] & 779.20 &  192.44\\
    Total Costs [€] & 2771.64 & 1247.67\\
    \hline
  \end{tabular}
\end{table}
The decreasing cost pattern breaks down for RC activation at 19:00 $D$ for all fleet sizes. For these cases, typical overloadings between 16:00 and 18:00 could only be addressed with CLCs, thereby giving the SO less flexibility to apply redispatch when beneficial. For $V=25.000$, we also see increased costs and slightly less activation of RC at 15:00$D$. For that situation, the trading window for finding buy volume for peak hours 17:00 and 18:00 $D$ is relatively short, increasing the risk of not finding enough counter volume in the redispatch market for some days in the test set. Redispatch activation at 11:00 $D$ is the best option for the large fleet.

\subsubsection{Policy Effectiveness and the Role of Risk Parameter $\epsilon$}
\begin{figure}[!t]
\centering
\includegraphics[width=0.85\linewidth]{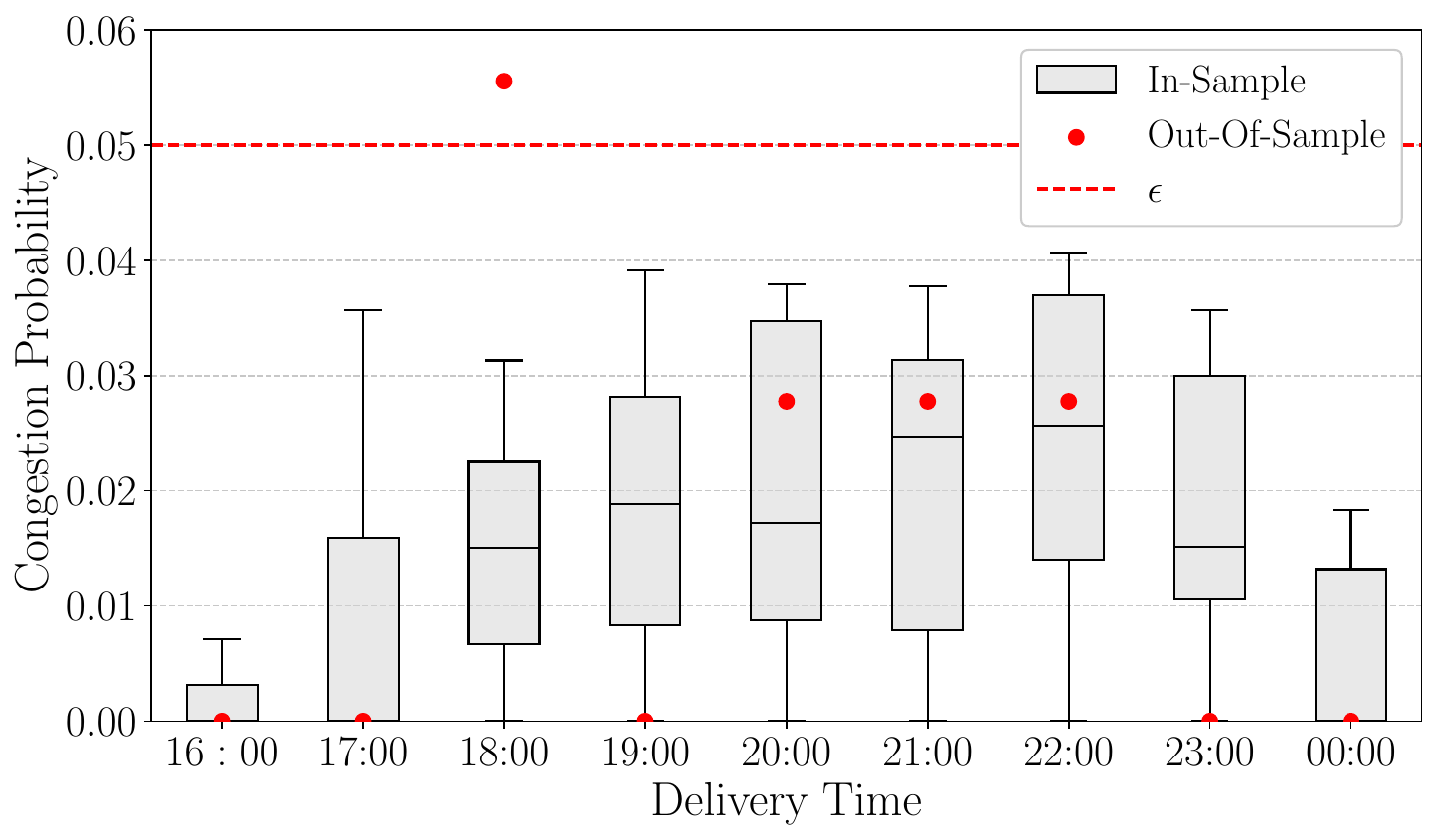}
\caption{In-sample and out-of-sample congestion probabilities for the 36 days in the test set for various delivery times. The value of $\epsilon=0.05$ denotes the accepted congestion probability set in the CVAR constraints \eqref{eq: CVAR a}-\eqref{eq: CVAR b}. The simulated EV fleet is of size $V=25.000$ and redispatch activation takes place at 11:00 $D$.}
\label{fig: probability}
\end{figure}

To investigate the effectiveness of the calculated policies, Fig. \ref{fig: probability} presents the congestion probabilities for various times on day $D$ for the most interesting large fleet size $V=25.000$ and the best redispatch activation moment at 11:00 $D$. The in-sample probabilities are presented as boxplots, as a probability can be calculated for all 36 days in the test set. If we apply the 36 policies on the realized EV and redispatch market data, we get 36 times congestion or not for every $d$. This relative frequency of congestion is also presented in Fig. \ref{fig: probability} with the red dots. We observe that the in-sample congestion probability is typically significantly lower than the value of $\epsilon=0.05$. This is due to the conservative CVAR approximation used to approximate the chance constraint. We also see that only for 18:00, the relative frequency of congestion exceeds $\epsilon$ as two out of the 36 test days showed congestion for that time. For one of the two days, the EV load was severely underestimated by the RF models. For the other, the realized redispatch counter volume was smaller than expected. For fleet sizes of $V=2.500$ and $10.000$, the out-of-sample congestion frequency did not exceed $\epsilon$ (data not shown).

\begin{figure}[!t]
\centering
\includegraphics[width=0.85\linewidth]{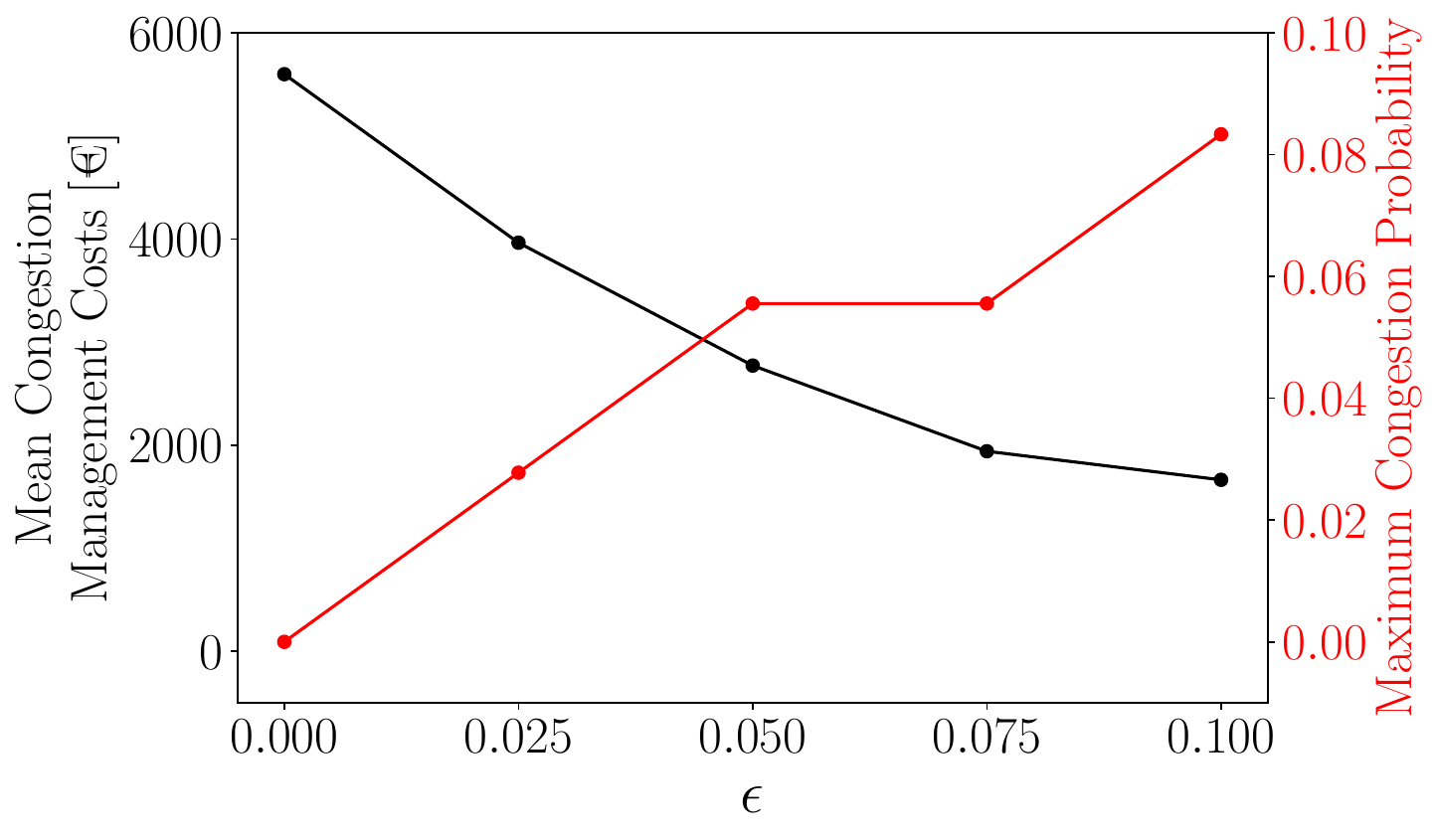}
\caption{The mean congestion management costs and maximum out-of-sample congestion probability over the day for various values of the accepted congestion probability $\epsilon$ set in the CVAR constraints \eqref{eq: CVAR a}-\eqref{eq: CVAR b}. The simulated EV fleet is of size $V=25.000$ and redispatch activation takes place at 11:00 $D$.}
\label{fig: epsilon}
\end{figure}

The low in-sample congestion probabilities indicate that the conservative CVaR-based approximation of the chance constraint may be overly restrictive. This opens up the possibility of reducing costs by selecting a higher value for $\epsilon$ in CVAR constraints~\eqref{eq: CVAR a}–\eqref{eq: CVAR b}. Although the original chance constraint targets a congestion probability of at most $5\%$, increasing $\epsilon$ beyond $0.05$ could partially offset the conservativeness introduced by the CVaR formulation. Figure~\ref{fig: epsilon} shows the mean congestion management costs and the maximum out-of-sample congestion probability across the day for various values of $\epsilon$, with fleet size $V = 25.000$ and redispatch activation at 11:00~$D$. As expected, relaxing the constraint by increasing $\epsilon$ leads to decreasing costs and increasing congestion probabilities. This way, reducing the out-of-sample congestion probability below 0.05 by decreasing $\epsilon$ to e.g. 0.025 can come with significant costs. On the other hand, increasing $\epsilon$ from $0.05$ to $0.075$ reduces costs without increasing the out-of-sample congestion probability. In other words, allowing more in-sample violations under the CVaR constraint does not necessarily lead to more congestion when the policy is applied out-of-sample.

\section{Conclusion}
\label{sec: conclusion}
This paper presents optimal congestion management with capacity limitation contracts (CLCs) and redispatch contracts (RCs) as a risk-aware resource allocation problem. Scenario generation for both EV fleet dynamics and the state of the national redispatch market (GOPACS) are developed based on real charging and market data.

The model seems to be conservative when calculating the in-sample congestion probabilities due to the CVAR relaxation of the chance constraints. The out-of-sample congestion probability was either below or close to the set target of 5.0\%. The optimal policies typically adopt a mix of the two instruments, hedging the risks associated with individual instruments. The RC was relatively expensive per unit due to frequent negative buy prices in the redispatch market, but the RC still proved effective if the congestion was relatively small (but larger than the minimum redispatch bid size) or if the prognoses communicated by the CSP underestimated the congestion. More generally, higher uncertainty about EV charging increased the use of RC intraday to correct forecasting errors at the CLC decision stage. In most cases, later RC activation was beneficial, as long as it happened before peak charging times after 16:00. This trend broke down for the largest fleet size considered, consisting of 25.000 cars. For these larger congestion volumes, it was not always certain that enough counter volume would be available in the redispatch market, reflecting the market liquidity risk associated with the RCs. Activating RCs late increased this risk, as it reduced the trading window for the redispatch products. For this fleet size, activating redispatch around 11:00 on the day of delivery seemed to strike the right balance. The results demonstrate the importance of various sources of uncertainty on trade-offs between congestion management instruments. Combining multiple instruments can be beneficial to hedge risks associated with individual instruments, but an understanding of how various conditions impact the risk profiles is vital for system operators for effective and efficient congestion management.

\section*{Acknowledgments}
The authors would like to thank the nominated electricity market operator ETPA for making available the anonymized limit order book dataset of GOPACS orders submitted through their platform.

 
%

\bibliographystyle{IEEEtran} 
\bibliography{bibliography}

\newpage

\vfill

\end{document}